\documentclass[useAMS,usegraphicx,usenatbib]{mn2e}

\usepackage{amssymb}
 \long\def\symbolfootnote[#1]#2{\begingroup%
 \def\thefootnote{\fnsymbol{footnote}}\footnote[#1]{#2}\endgroup}

\usepackage{subfigure}

\title[Protocluster Galaxies Around $z = 2.2$ Radio Galaxies]{Early Quenching of Massive Protocluster Galaxies Around $z=2.2$ Radio Galaxies }
\author[K. Husband et al.]{K. Husband$^1$, M.~N. Bremer$^1$\thanks{Email: m.bremer@bristol.ac.uk}, J.~P. Stott$^{2,3}$, D.~N.~A. Murphy$^4$  \\
$^1$H.H.~Wills Physics Laboratory, University of Bristol, Tyndall Avenue, Bristol, BS8 1TL, UK. \\
$^2$Sub-department of Astrophysics, Department of Physics, University of Oxford, Denys Wilkinson Building, Keble Road, Oxford OX1 3RH, UK. \\
$^3$Institute for Computational Cosmology, Durham University, South Road, Durham DH1 3LE, UK. \\
$^4$Instituto de Astrof\'isica, Pontificia Universidad Cat\'olica de Chile, Avenida Vicu\~na Mackenna 4860, Santiago, Chile. \\
}

\begin{document}
\maketitle
\begin{abstract}
Radio galaxies are among the most massive galaxies in the high
redshift universe and are known to often lie in protocluster
environments. We have studied the fields of seven $z=2.2$ radio
galaxies with HAWK-I narrow-band and broad-band imaging in order to
map out their environment using H$\alpha$ emitters (HAEs).  The
results are compared to the blank field HAE survey HiZELS. All of
the radio galaxy fields are overdense  in HAEs
relative to a typical HiZELS field of the same area and four of the
seven are richer than all except one of 65 essentially random HiZELS
subfields of the same size. The star formation rates of the massive
HAEs are lower than those necessary to have formed their stellar
population in the preceding Gyr - indicating that these galaxies are
likely to have formed the bulk of their stars at higher redshifts, and
are starting to quench.
\end{abstract}


\section{Introduction}
Overdensities of galaxies that are expected to be the progenitors of
local massive galaxy clusters have been found around high redshift
radio galaxies \citep[e.g.][]{Venemans04,Overzier06,Hatch11,Kuiper11}
and quasars \citep[e.g.][]{Venemans07,Kim09,Utsumi10,Husband13}. These
protoclusters are generally discovered via their star-forming
population; in part because it is easier to get confirming
spectroscopy of actively star-forming galaxies that contain emission
lines, unlike passive galaxies, and in part because studies of low and
intermediate redshift clusters indicate that the majority of stars in 
cluster galaxies formed at $z>2$ \citep[e.g.][]{Ellis97,Tran07}. This
 rapid growth in clusters at high redshift contrasts to that in low
redshift clusters where star formation is suppressed relative to the
field. The redshift range over which their galaxy population becomes
red and dead  can be determined using a large sample of protoclusters
selected through a range of techniques in order to minimize selection
biases.

An efficient way of finding protoclusters at $z>2$ appears to be
through targeted searches around radio galaxies and quasars.  The
  growth of galaxies is likely linked to the growth of their central
  black holes, and consequently AGN are expected to reside in
  protoclusters \citep{Smail03,
    Lehmer09,Digby-North10,Matsuda11}. This and the fact that
powerful radio galaxies are generally among the most massive galaxies
at any epoch \citep{deBreuck02, Seymour07} makes them ideal
objects for targeted protocluster searches. There is already a
significant body of work exploring radio galaxy environments through
H$\alpha$ emission such as that by \citet{Hatch11} using HAWK-I and
ISAAC on the VLT and the Mahalo (`Mapping HAlpha and Lines of Oxygen
with Subaru') project with Subaru \citep{Kodama13, Shimakawa14} among
others \citep[e.g.][]{Cooke14}. These H$\alpha$ studies have often
targeted known protoclusters discovered by other means (such as
overdensities of red galaxies or BzKs) and may well be subject to
publication bias where only the most overdense regions are followed up
or published, giving little clue to the fraction of radio galaxies
that reside in star-forming overdensities.

Powerful radio sources themselves significantly influence the
evolution of galaxies within their host dark matter halo. Radio jets
are known to stop gas cooling through the kinetic mode of feedback on
galactic scales \citep{McNamara07, Cattaneo09} and powerful radio
galaxies at high redshift, whose jets can extend over 100s of kpc, may
also affect intra-group gas \citep{Fabian12}. Outflows from $z\sim2$
radio galaxies may be observational evidence for radio jets
interacting with the early intra-group or intra-cluster medium
\citep{Nesvadba06,Nesvadba08}. Such AGN feedback is essential in
simulations to reproduce the observed anti-hierarchical growth and
local galaxy luminosity function \citep[e.g.][]{Bower06}. AGN feedback
on extragalactic scales may increase the entropy and pressure of the
gas in the local environment of massive galaxies cutting off the
supply of cold gas, which would otherwise accrete on to the galaxies
fuelling star formation, and resulting in relatively quiescent member
galaxies relatively early on \citep{Hatch14}.

In this work we have explored the $\sim$12 co-moving Mpc scale
environment of seven $z=2.2$ radio galaxies with VLT/HAWK-I using
H$\alpha$ emitters (HAEs) selected through narrow-band imaging in
order to study galaxy clustering around such objects. The
seven radio galaxies were selected purely on their spectroscopic
redshift (falling within the range of the HAWK-I narrow-band filters)
and availability from Chile on the dates of observations. They all
have radio luminosities greater than 1 $\times$ 10$^{26}$ W
  Hz$^{-1}$ at 4.7-4.85 GHz observed. Selecting H$\alpha$
emitters results in a relatively clean sample of galaxies within a
narrow redshift range ($\Delta z=0.05$) as H$\alpha$ is less affected
by dust extinction (or metallicity) compared to other strong lines
\citep{Koyama13a}. We use the same method as the HiZELS survey
\citep{Sobral13} to select HAEs in order to have a field galaxy
comparison sample.

A $\Lambda$CDM cosmology with H$_0=69.6$ km s$^{-1}$ Mpc$^{-1}$,
$\Omega_M=0.286$ and $\Omega_{\Lambda}=0.714$ \citep{Bennett14} is
used throughout and all the magnitudes quoted are in the AB system
\citep{Oke83}.

\section{Data}
\subsection{Imaging and Data Reduction}
The seven radio galaxy fields were imaged with HAWK-I (High Acuity
Wide field K-band Imager) on the VLT in Oct/Nov 2012 and Jan/Feb/Mar
2013 with the $J$ filter,  short K ($Ks$) filter and a
narrow-band filter centered on the wavelength of H$\alpha$ from the
radio galaxy (NB2090, H2 or Br$\gamma$). HAWK-I has a field of view of
7.5 by 7.5 arcmin$^2$ or 12.2 by 12.2 co-moving Mpc$^2$ at these
redshifts ($z=2.23$).  The average exposure time was 0.62, 0.71 and
3.7 hours in $J$, $Ks$ and the narrow-band (NB) reaching 2$\sigma$
depths of 22.9, 23.0 and 22.4 on average respectively (see Table
\ref{imag_summary}).

The radio galaxies were  selected in a unbiased way from a narrow
  redshift range between $2.198<z<2.294$ to match the available NB
  filters. They lie over a range of RAs convenient for scheduling.
Only the environment of MRC0200+015 has been studied before - it was
found to be overdense in HAEs by \citet{vanderWerf00} and
\citet{Matsuda11} but our new observations are $\sim$1 magnitude
deeper.

The data was reduced by first subtracting a dark frame from the images
and then flat fielding with an averaged, normalised twilight flat
field. The images were then normalised and median combined together
without offsets to make another flat field that was applied to all
images to remove any remaining sky residuals. Finally, the images were
combined with offsets, cosmic ray rejection (using sigma clipping) and
a bad pixel mask in order to deal with the chip gaps. The images were
calibrated using unsaturated and cleanly extracted 2MASS objects in
the fields.  The magnitudes of the objects were extracted in 2 arcsec
diameter apertures using SExtractor \citep{Bertin96}.

\begin{table*}
   \centering
    \begin{tabular}[t]{c c c c c c c c c c}
    \hline
    Field & RA  & Dec.  & Redshift & L$_{4.8 GHz}$  & NB filter & NB Exposure  & K Exposure & NB Seeing  & K Seeing  \\
     &  & & &  /10$^{26}$ W Hz$^{-1}$ & & /h (2$\sigma$ AB) & /h (2$\sigma$ AB) & /arcsec &  /arcsec \\
    \hline
   MRC 0200+015  & 02:02:42.9 & +01:49:10 & 2.229 & 21.1 & H2        & 3.33 (22.5) & 0.66 (22.9) & 0.59 & 0.68\\
   NVSS J015640  & 01:56:40.4 & -33:25:33 & 2.198 & 42.4 & NB2090    & 3.33 (22.2) & 0.66 (23.4) & 0.60 & 0.64\\
   PMN J0340-6507& 03:40:44.9 & -65:07:07 & 2.289 & 33.8 & Br$\gamma$& 4.70 (22.4) & 0.66 (22.8) & 0.70 & 0.74\\
   NVSS J045226  & 04:52:26.6 & -17:37:53 & 2.256 & 9.6  & H2        & 4.00 (22.5) & 0.66 (22.9) & 0.55 & 0.72\\
   NVSS J094748  & 09:47:48.4 & -20:48:36 & 2.294 & 4.0  & Br$\gamma$& 3.33 (22.4) & 0.66 (22.6) & 0.54 & 0.77\\
   NVSS J100253  & 10:02:53.1 & +01:34:56 & 2.248 & 1.6  & H2        & 3.33 (22.7) & 0.66 (23.0) & 0.58 & 0.59\\
   MRC 1113-178  & 11:16:14.5 & -18:06:22 & 2.239 & 30.0 & H2        & 3.62 (22.2) & 0.66 (23.3) & 0.65 & 0.62\\
    \hline
    \end{tabular}
    \caption{ A summary of the HAWK-I imaging. The radio luminosities
      are from observations at 4.7 or 4.85 GHz. The NB filters
        used in this work were NB2090 ($\lambda_{mean}=20954$ \AA;
        covering H$\alpha$ between $z=2.178-2.207$), H2
        ($\lambda_{mean}=21248$ \AA; covering H$\alpha$ between
        $z=2.215-2.260$) and Br$\gamma$ ($\lambda_{mean}=21643$ \AA;
        covering H$\alpha$ between $z=2.275-2.321$). The K filter used
        was the $Ks$ filter (central wavelength = 21323 \AA, FWHM =
        3150 \AA). }
    \label{imag_summary}
\end{table*}

\subsection{H$\alpha$ Emitter Selection}
\begin{figure}
\centering
\subfigure{\includegraphics[width=0.99\columnwidth]{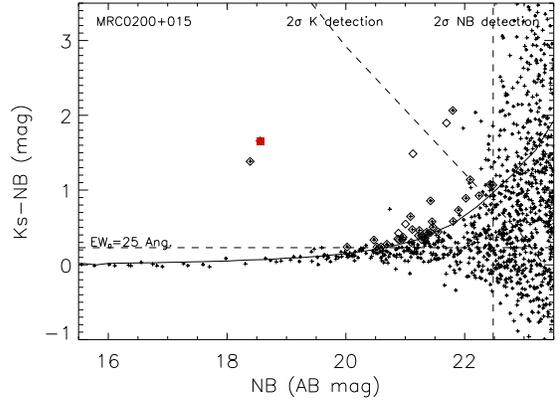}}
\caption{The $Ks-$narrow-band ($NB$) vs.~$NB$ colour magnitude diagram for
  the MRC0200+015 radio galaxy field. The red square indicates the
  radio galaxy and the dashed lines show the 2$\sigma$ limits on the
  imaging. Also shown is the equivalent width limit and the line of
  three times the average observational error. The HAEs that were
  selected after visual inspection are highlighted by diamonds. }
\label{MRC0200+015}
\end{figure}

\begin{figure}
\centering
\subfigure{\includegraphics[width=0.95\columnwidth]{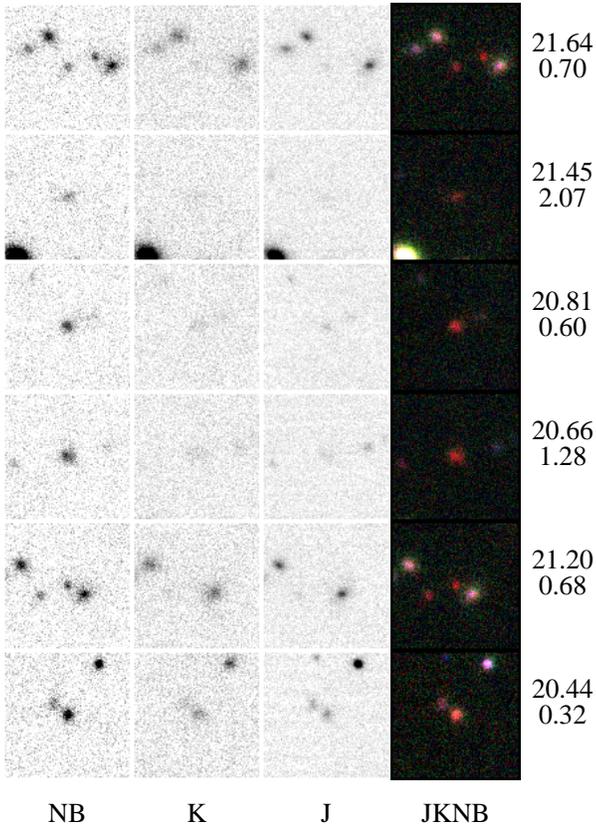}}
\caption{The narrow-band ($NB$), $Ks$, $J$ and three colour combined
  images of some of the H$\alpha$ emitters selected (all taken from
  the NVSS J094748 radio galaxy field). The images are 10 by 10 arcsec
  across and the numbers on the right hand side are the NB AB
  magnitudes and the $Ks-$NB colours of the HAEs. }
\label{HAEexample}
\end{figure}

We selected HAEs from a $Ks$-NB vs.~$Ks$ colour magnitude diagram (see
for example the colour magnitude diagram for MRC0200+015 in
Fig.~\ref{MRC0200+015}) in a similar manner to
\citet{Sobral13}. Specifically in this work HAEs are defined as
galaxies with a $Ks$-NB colour greater than  3$\Sigma$ (where
  $\Sigma$ is the combined average error on the NB and $Ks$ band
  magnitudes at the NB magnitude), a NB magnitude brighter than the
  2$\sigma$ NB limiting magnitude, and a rest-frame narrow band
equivalent width greater than 25 \AA. All of the HAEs were
individually checked to confirm that their sizes and morphologies were
consistent with $z\sim2$ galaxies rather than stars or
artefacts. Fig.~\ref{HAEexample} shows some of the selected HAEs.

\subsection{H$\alpha$ Star Formation Rates and Equivalent Widths}
The star formation rates (SFRs) and equivalent widths (EWs) of the
HAEs were calculated from the NB and $Ks$ magnitudes, via the
continuum flux density per Angstrom, $f_{Kc}$, and H$\alpha$ flux,
$f_{H\alpha}$, using the following equations \citep[see for
  example][]{Cooke14}:

\begin{equation}
f_{Kc} = \frac{w_K10^{(-m_K-48.6)/2.5}-w_{NB}10^{(-m_{NB}-48.6)/2.5}}{w_K-w_{NB}}
\end{equation}
\begin{equation}
f_{H\alpha} = w_{NB}(10^{(-m_{NB}-48.6)/2.5}-f_{Kc})
\end{equation}
\begin{equation}
EW=\frac{f_{H\alpha}}{f_{Kc}(1+z)}
\end{equation}
\begin{equation}
SFR=4\pi d^2 f_{H\alpha}\times 4.39\times10^{-42}\: M_{\odot}\, yr^{-1}
\end{equation}

where $w_K$ and $w_{NB}$ are the effective widths of the $Ks$ and NB
filters, $m_K$ and $m_{NB}$ are the $Ks$ and NB AB magnitudes of the
HAEs, $f_{H\alpha}$ is the flux of H$\alpha$ in erg s$^{-1}$
cm$^{-2}$, $d$ is the co-moving radial distance in centimeters and $z$
is the redshift of the HAEs, which is assumed to be the same as the
radio galaxy.  Equation 4 assumes that all of the photoionization is
by young stars and not active galactic nuclei (AGN). If AGN are
present then the estimates of SFR will be too high. However previous
follow-up of HAEs with X-ray observations and rest-frame optical
spectroscopy, in both clustered and non-clustered fields, indicates
only a low ($<$10 per cent) fraction of HAEs contain AGN \citep[see][]{Sobral13,Koyama13a,Stott13, Hatch11}. Hence, we do not expect this
to significantly affect our results.

\subsection{HAE Masses}
The masses of the HAEs were estimated from the observed
$Ks$-band magnitudes,  using a mass-to-light ratio with an
  additional $J-Ks$ colour term to take into account different star
  formation histories, following the method of
\citet{Koyama13a}. Specifically:
\begin{equation}
\log(M_{*}/10^{11} M_{\odot})=-0.4(Ks-22.24)+\Delta \log M
\end{equation}
where
\begin{equation}
\Delta \log M=0.14-0.9\exp[-1.23(J-Ks)]
\end{equation}
and $M_{*}$ is the stellar mass, $J$  and $Ks$ are  AB
magnitudes, and assuming a Salpeter IMF. We then convert these to the equivalent Chabrier masses for consistency with equation (4).
  \citet{Koyama13a} note that this ``one-colour method'' agrees well
  with a full SED fitting method (with $\sim 0.3$ dex scatter) over a
  wide range of luminosities (over nearly 3 magnitudes).

Again if a HAE contains an optically bright AGN then the estimate of
its mass will be too high, but we do not expect a large AGN fraction
(see previous and next section) and so this should not
significantly affect our results.

\subsection{Contamination}
The final sample of HAEs could be contaminated by emission line
galaxies such as [OIII] emitters at $z\sim3$ or Pa series emitters at
lower redshifts. The higher redshift interlopers are likely very rare
\citep[for example only 1 of 55 HAEs satisfied a $z\sim3$ LBG
  selection in][]{Geach08} and broad-band selections such as the BzK
selection \citep{Daddi04} can remove the lower redshift
interlopers. In previous studies the majority ($>90$ per cent) of HAEs
were found to lie within the BzK selection
\citep{Sobral13,Koyama13a}. Due to this and the paucity of deep
ancillary multi-wavelength data in these fields we do not apply
additional broad-band selections (such as BzK) here as
\citet{Sobral13} has shown it to be unnecessary. As the probability
for any one HAE detection to be a contaminant is small ($\lesssim 10$
per cent), the probability for a group of contaminant galaxies to
align with the radio galaxy is very small, and hence we believe
contaminants do not significantly effect this work.

\section{Results \& Discussion}
\subsection{Radio Galaxy Environments}
\begin{figure*}
\centering
\subfigure{\includegraphics[width=0.65\columnwidth]{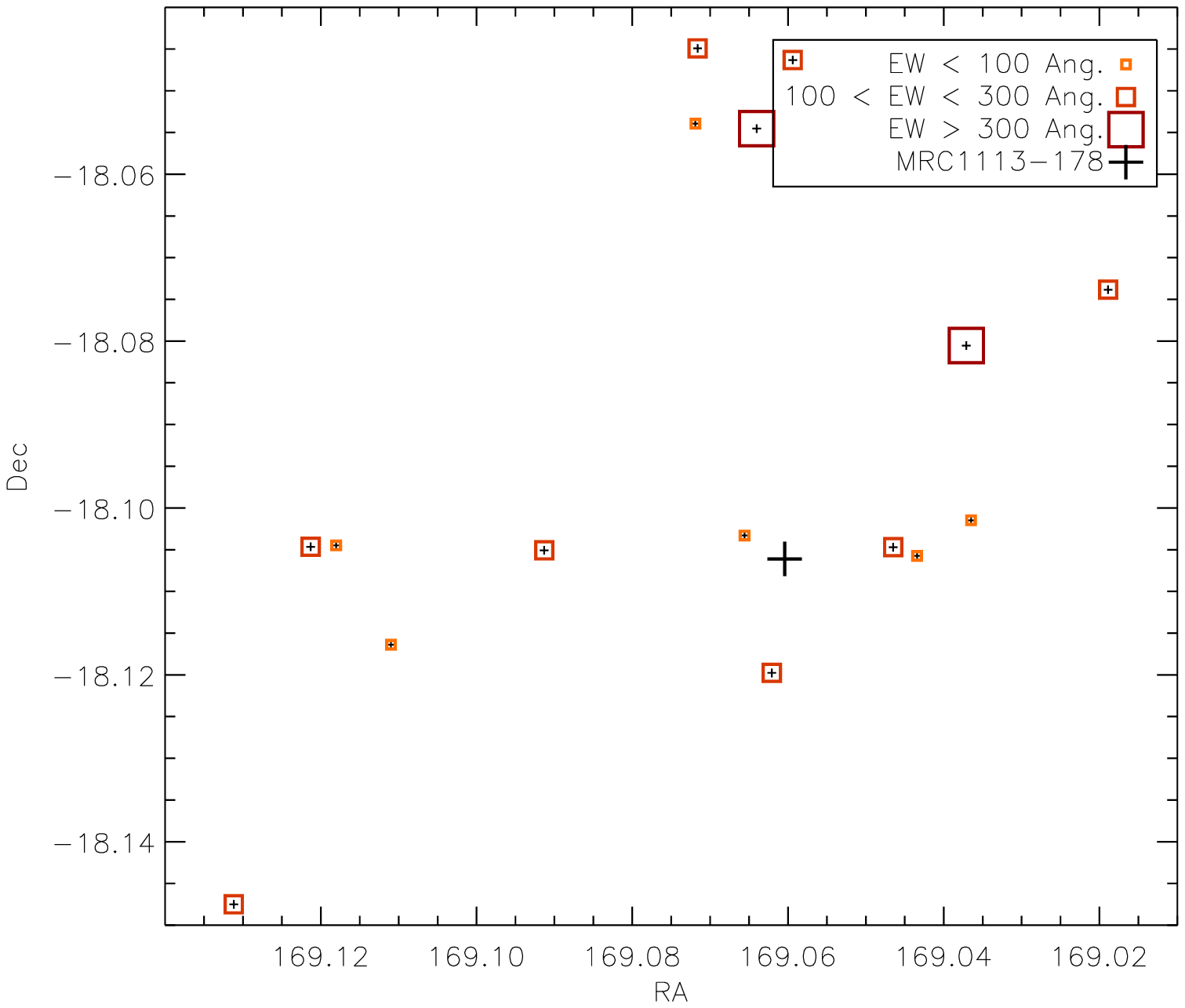}}
\subfigure{\includegraphics[width=0.65\columnwidth]{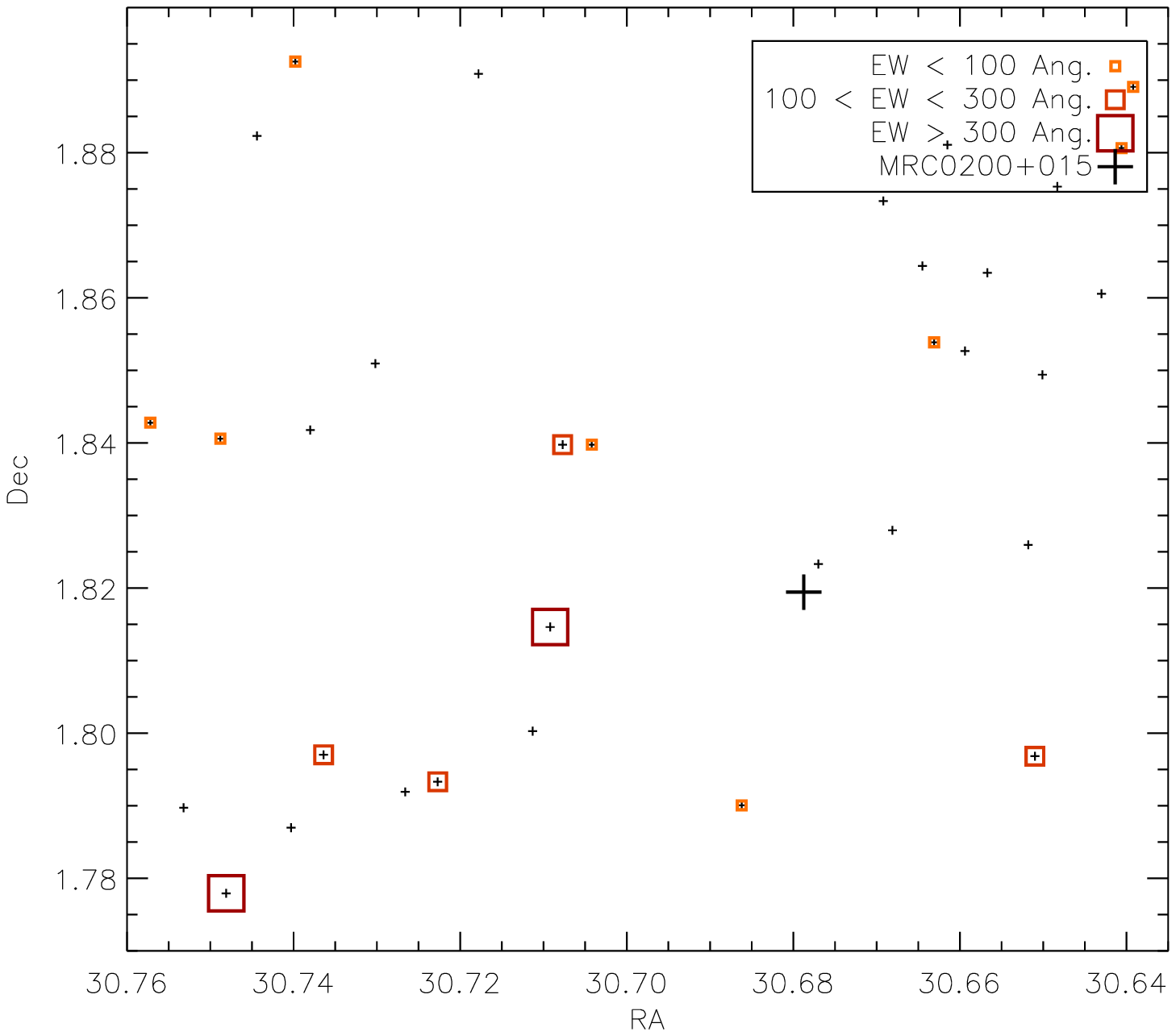}}
\subfigure{\includegraphics[width=0.65\columnwidth]{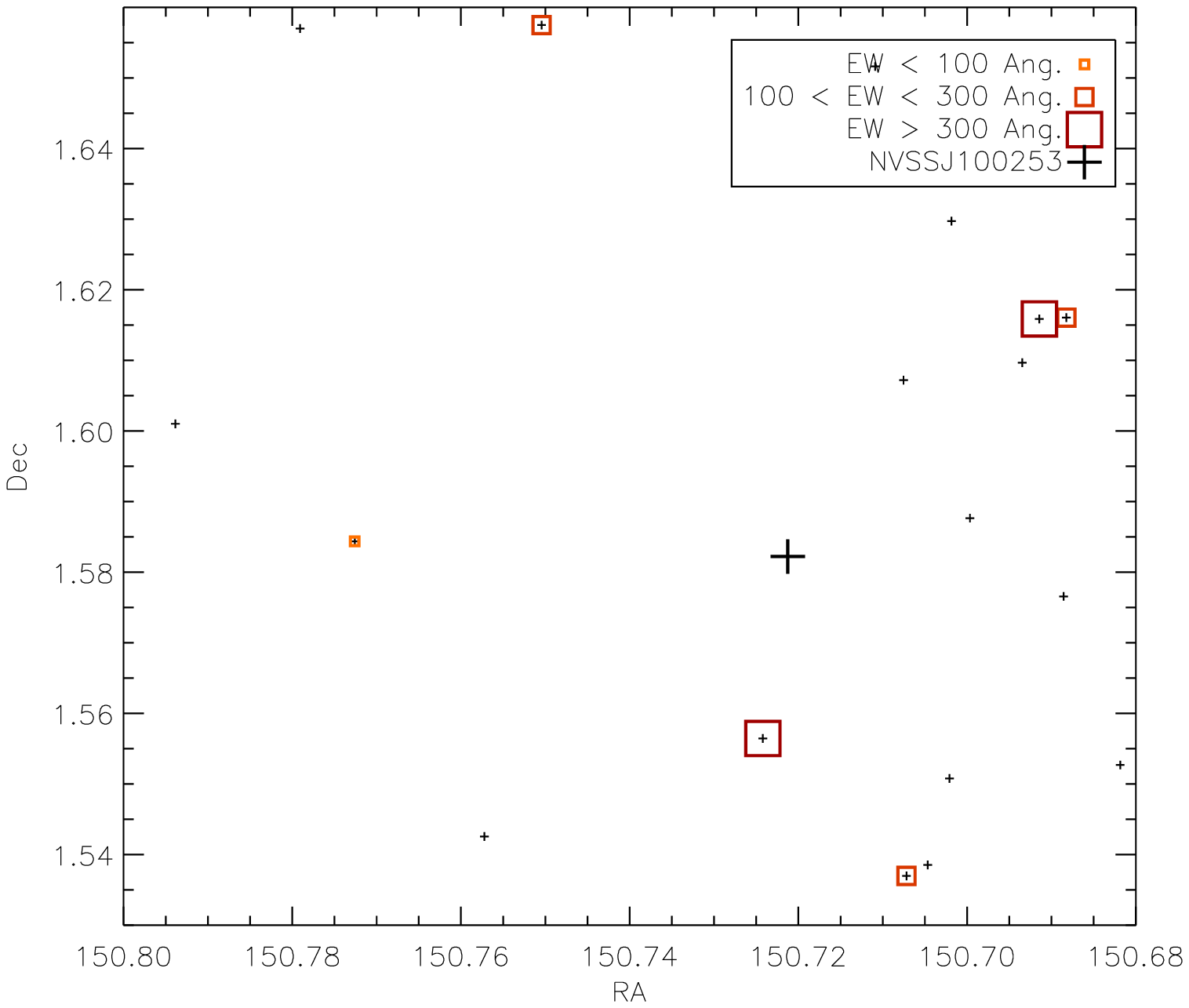}}
\caption{The distribution of H$\alpha$ emitters that meet the NVSS
  J094748 selection criteria in the two richest (left and middle) and
  the poorest (right) of the radio galaxy fields. The size and colour
  of the point indicates the equivalent width of the HAE. The small
  crosses indicate HAEs that do not meet the NVSS J094748 selection
  criteria. At this redshift 0.01 degrees corresponds to 0.97
  co-moving Mpc. }
\label{pos_together_EW}
\end{figure*}

The positions of the HAEs in the two richest and the poorest radio
galaxy fields are shown in Fig.~\ref{pos_together_EW}. The number of
HAEs found in each field is detailed in Table \ref{passive}. As the
image depth varies between fields, the number of HAEs in each field
using an identical selection is also shown in the table having applied
the selection function of the shallowest field, NVSS J094748, and
corrected the NB magnitudes to those expected if the same narrow-band
filter was used as the NVSS J094748 field.

In order to understand the significance of any clustering in the radio
galaxy fields, we can compare the number of HAEs in each field
to those derived from the much larger area HiZELS observations carried
out with UKIRT \citep{Sobral13}. HiZELS imaged 2.3 deg$^{2}$ of COSMOS
and UDS to a similar depth as the radio galaxy fields and selected
HAEs at the same redshift with a similar criteria and the same
equivalent width limit as this work.  However, HiZELS uses a smaller
width narrow-band filter and hence probes only $\sim$0.7 times the
volume per unit area in comparison to the observations of all radio
galaxy fields except that of NVSS J015640. The difference in filter
widths also results in a different relationship between $K_s-NB$
colour and line equivalent width. In all of the following we  scale
the HiZELS-derived numbers to the volume and equivalent-width
sensitivity of our data.

\begin{table*}
   \centering
    \begin{tabular}[t]{c c c c c c}
    \hline
    Field & No.~of HAEs & No. of HAEs &  No.~of Bright HAEs  & Overdensity$^2$, $\rho_g$  & Mass at $z=0$ \\
      & & to same limit$^1$ & (L$_{H\alpha} >10^{43}$ erg s$^{-1}$) & (excess)&  /$10^{14}$ M$_{\odot}$ \\
        \hline
    MRC 0200+015   & 39 & 14 & 5  & $3.2\pm1.1$ &  $12\pm4.3$ \\
    NVSS J015640 & 10 & 7.8 (5) & 3 (2) & $1.4\pm0.9$ & $3.2\pm2.0$ \\
    PMN J0340-6507 & 24 &  8 & 5  & $1.4\pm0.9$ & $5.4\pm3.2$ \\
    NVSS J045226   & 32 & 12 & 5  & $2.6\pm1.1$ & $8.4\pm3.4$  \\
    NVSS J094748   & 13 & 13 & 4  & $2.9\pm1.1$ & $6.3\pm2.4$ \\
    NVSS J100253   & 18 &  5 & 1  & $0.8\pm0.7$ & $2.7\pm2.5$  \\
    MRC 1113-178   & 16 & 16 & 10 & $3.9\pm1.2$ & $12\pm3.9$  \\
    Radio Galaxy Mean &  23.4 & 11.0 &  4.6 & 2.3 & 7.2  \\
    HiZELS &  na & 3.3 (2.4) & 0.6 (0.4) & na & na \\
   \hline
    \end{tabular}
    \caption{A summary of the number of galaxies and relative
      overdensity detected in each field. Brackets denote the raw
        number of galaxies measured in the fields, NVSS J015640 and
        HiZELS, that have been corrected for the narrower filter
        widths. $^1$The selection of the shallowest field, NVSS
        J094748, is applied to all fields so a direct comparison can
        be made. $^2$Note that the overdensity is calculated to be the
        number of galaxies in excess of the background
        i.e.~$\rho_g=(\rho_{rg}-\rho_{bkg})/\rho_{bkg}$ where $\rho_{rg}$ is
        the surface density of galaxies in the protocluster fields and
        $\rho_{bkg}$ is the surface density of background galaxies
        calculated from the HiZELS survey using the same HAE selection
        as the NVSS J094748 field \citep{Rigby14}. The upper limit to
        the expected eventual mass of a system at $z=0$ are calculated
        using the matter overdensity method discussed in the text The
        errors on these masses are calculated taking account of the
        statistical uncertainty on the number of excess HAEs measured
        in each field. }
    \label{passive}
\end{table*}

 We explore the strength of clustering by a simple counts in cells
 analysis, each cell being the size of a HAWK-I field. We determine
 the number of line emitters that would meet our NVSS J094748
 selection criteria having taken into account the different width NB
 filters used in the two sets of observations. We place the cells onto
 the HiZELS data in two ways. Firstly, we simply divide the HiZELS
 surveys into 88 equal-area, non-overlapping squares or cells. As this
 does not take into account any intrinsic clustering in the $z=2.2$
 galaxy distribution, we secondly amend the positioning of each cell
 so that it is centred on a HAE, (to mimic the effect of the HAWK-I
 fields being centred on known $z=2.2$ galaxies) while ensuring the
 cells still do not overlap. This necessarily reduces the number of
 cells to 65 as the spatial distribution of HAEs does not allow for as
 efficient abutting of cells as simply splitting the entire survey
 area uniformly. In reality, the difference in the statistics derived
 from the two approaches is very similar with the mean number of
 sources per cell meeting our selection criteria increased by only
 $\sim 30$ per cent when they are centred on HAEs. We use the
 statistics derived from the second approach in the following
 analysis.

\begin{figure}
\centering
\includegraphics[width=0.95\columnwidth]{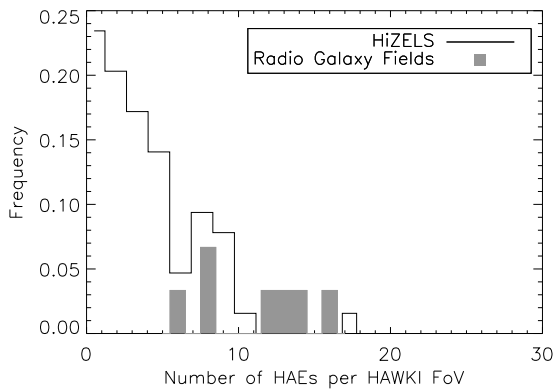}
\caption{The number of HAEs meeting our selection criteria around
    non-overlapping HAWK-I sized pointings centered on HAEs in the
    HiZELS fields (COSMOS+UDS). The number of HAEs around the radio
    galaxies are shown by grey-filled bins whose frequency is set to
    an arbitrary level. }
\label{density}
\end{figure}

The distribution of the  number of HAEs per HAWK-I field derived
  from HiZELS is shown in Fig.~\ref{density}.  The volume density of
HAEs derived from the HiZELS data translates to a mean surface density
of 3.3 per HAWK-I survey field.  As summarised in Table
  \ref{passive} the radio galaxy fields are on average  three
  times denser than the HiZELS survey fields and one field
(MRC1113-178) in particular contains nearly five times the number of
HAEs than the mean HiZELS value.  The highest density field out of the
65 in the HiZELS distribution (the cell with 17 HAEs) is contributed
by a single structure in one of the two HiZELS fields. This structure
has been discussed by \citet{Geach12} and is likely to turn into a
significant cluster at $z=0$. While it does not contain any radio
source of comparable luminosity to those studied here, it does contain
a quasar at the same redshift.

\begin{figure}
\centering
\includegraphics[width=0.95\columnwidth]{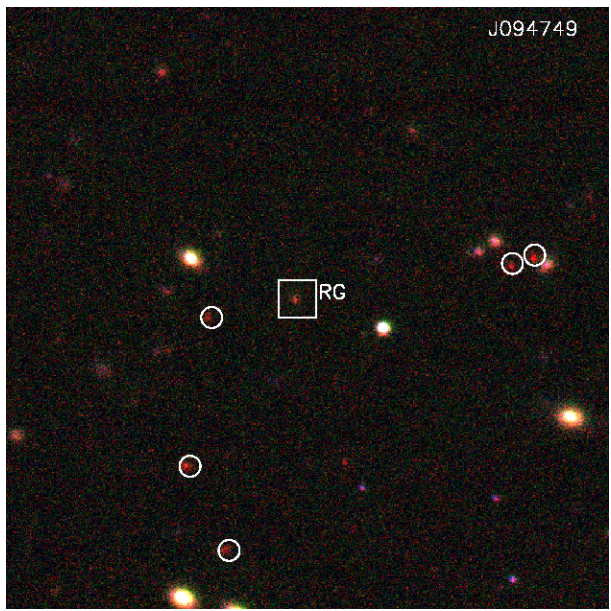}
\caption{The 1 by 1 arcmin$^2$ narrow-band image of the centre of the
  NVSS J094748 field (the online journal shows the three-colour,
  $JKsNB$, image). The radio galaxy is marked by a square and the
  redder HAEs that meet our selection criteria are circled. }
\label{NVSSJ09}
\end{figure}

Although the HAWK-I field of view is well-matched to the predicted
effective radius of protoclusters from the Millennium Simulation
\citep[ $\sim$ 6 co-moving Mpc;][]{Chiang13}, some sub-clustering is
expected particularly near the central massive object. Indeed, in some
fields the overdensity is much larger if we consider a smaller
scale. In particular, in the NVSS J094748 field the HAEs appear to
cluster around the radio galaxy (see Fig.~\ref{NVSSJ09}). In a 1
arcmin$^2$ area there are five HAEs plus the radio galaxy compared to
an expectation of $\sim$0.2 HAEs per arcmin$^{-2}$ from the HiZELS
survey - only one of the HAEs in HiZELS has more HAEs within a 1 by 1
arcmin$^2$ box centered on them when scaled to the same volume,
indicating that the radio galaxy is at the centre of a dense structure
that perhaps evolves into a massive galaxy by the present day.

Given these results, the typical radio galaxy field contain a clear
excess of star-forming galaxies relative to the survey fields in line
with the literature \citep[e.g.][]{Hatch11,Kuiper11}, but with
significant variations from field to field. We find 4 out of 7 (around
60 per cent) of the radio galaxy fields to be denser than 98 per cent
(and all of the radio galaxy fields to be denser than 80 per cent) of
similar sized regions in HiZELS at $z=2.23$ when scaled to the same
volume per unit area. This is a similar result to \citet{Venemans07}
who found that 6 out of 8 of the $z>2$ radio galaxies in their sample
were surrounded by an overdensity of Ly$\alpha$ emitters. However,
from Fig.~\ref{density} it is clear that, on average, radio galaxies
at $z=2$ do not lie in the most extreme ($> 5 \sigma$) overdensities -
the existence of the \citet{Geach12} system within the HiZELS fields
demonstrates this.

Assuming that these overdensities could develop into current-day group
and clusters, it is instructive to estimate the likely eventual masses
of these systems. This can be done by estimating the matter
overdensity they represent, which in turn is related to the galaxy
overdensity measured through the galaxy bias, b \citep[following
  e.g.][]{Venemans07}. For HAEs at this redshift, selected in a
similar manner to ours down to a SFR limit of 20 M$_{\odot}$
yr$^{-1}$, the bias is measured to be around 2.4 \citep{Geach12}. If
we assume that all the matter within the volume will collapse into a
cluster by the present day, then the final mass of the system is just
the volume  ($\sim$ 5000 co-moving Mpc$^3$) times the matter
overdensity times the critical density of the universe. This gives
$z=0$ masses of at most several times $10^{14}$ M$_{\odot}$ (see Table
\ref{passive}).  However, these values must be taken as upper limits
as it is improbable that everything within the volume probed will
collapse into the eventual structure. The mass of these systems can
also be estimated independently, by mapping their apparent number
density onto the current-day cluster mass function.  As there is at
most one system of similar or greater density to the most clustered
radio galaxy field in the 2.34 deg$^2$ of HiZELS, the number density
of such systems must be around or less than $1\times10^{-6}$
Mpc$^{-3}$ implying the eventual mass of the richest of the systems
studied here would be $\sim 5\times10^{14}$ M$_{\odot}$ using the
\citet{Tinker08} $z=0$ halo mass function of clusters. Given the
inevitable scatter in the mass growth of individual structures between
$z\sim 2$ and today, both mass estimates are consistent and imply the
systems have the potential to become systems characterised as rich
groups or moderate-mass clusters today.

There have been numerous previous studies of radio galaxy environments
at z $\sim2$ using HAEs to map out the local galaxy densities
\citep[e.g.][]{Kurk04,Hatch11,Hayashi12,Koyama13b,Cooke14}. The
estimated final masses of these systems are again generally around a
few times $10^{14}$ M$_{\odot}$ using the galaxy bias prescription,
with the exception of the protocluster around the Spiderweb galaxy
whose eventual mass is estimated to be nearly $10^{15}$ M$_{\odot}$
\citep[both using this method and other methods based on additional
  data such as spectroscopic velocities and X-ray observations
  detailed in][]{Shimakawa14}.

\subsection{Luminosity and Mass Functions}
\begin{figure}
\centering
\includegraphics[width=0.95\columnwidth]{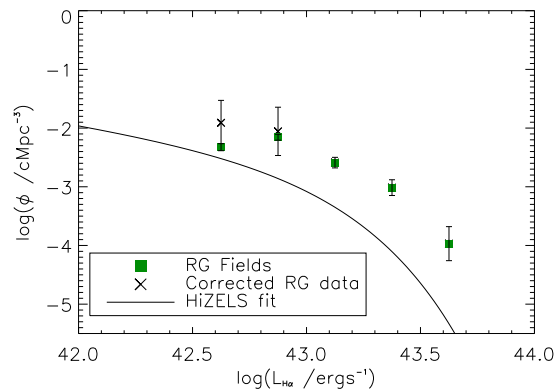}
\caption{H$\alpha$ luminosity function for the radio galaxy fields
  with the $z=2.23$ fit from the HiZELS survey work \citep{Sobral13}
  shown as a black line. The green filled squares are the observed
  luminosity densities and crosses are the values corrected for
  incompleteness following the prescription of \citet{Sobral13} which
  appreciably affects only the two low lowest luminosity bins. The
  error bars are generated as described in the text.}
\label{Lumfn_com}
\end{figure}

The H$\alpha$ luminosity function is shown in Fig.~\ref{Lumfn_com}
along with the luminosity function of field galaxies from
\citet{Sobral12}. The H$\alpha$ fluxes have been corrected for [NII]
emission that is likely to fall into the narrow-band filter. This was
carried out using an empirically derived relation between the ratio of
[NII] to H$\alpha$ and the sum of their equivalent widths taken from
\citet{Sobral12}. The median value of the [NII]/(H$\alpha$+[NII])
ratio is 0.16. In addition, \citet{Sobral09} have demonstrated
  that the wavelength response of similar filters are sufficiently
  close to a ``top-hat" profile that any difference has a minimal
  effect on the calculated luminosity function. Hence, we do not
  correct for the filter profile.

Numerous studies have  indicated that HAEs are dust extincted by
around $A_{H\alpha}= 1.0$ magnitude and that the amount of dust
extinction does not significantly change with luminosity
\citep[e.g.][]{Garn10,Sobral12}, although there is some evidence that the
  amount of dust correction may slightly depend on mass
  \citep[e.g.][see below]{Shimakawa15}. We follow
\citet{Sobral13} and apply one magnitude of dust extinction to all of
our HAEs. This will increase the SFRs inferred for the objects using
the relation given in section 2.4 but will not affect the masses as
this relation is based on the observed Ks magnitude.

In order to calculate the errors on the luminosity function, we
performed a Monte Carlo simulation whereby each HAE candidate was
simulated a thousand times with the $Ks$ band and $NB$ magnitudes
taken from a Gaussian distribution centered on the observed magnitudes
with a width equal to the error on the photometry. Assuming Poisson
statistics, the error on a particular luminosity bin is the square
root of the mean number of simulated HAEs falling within that bin
\citep[see below; this follows the method of][]{Sobral12}. The lowest
luminosity bins are affected by incompleteness and we correct for this
using the prescription of \citet{Sobral13}.

The H$\alpha$ luminosity function shows an excess of HAEs in the radio
galaxy fields compared to HiZELS. This is not due to the radio
galaxies themselves, which are excluded from the luminosity function
to avoid biasing the results as the rarity of radio galaxies means
they are not likely to contribute significantly to the HiZELS results.
  
The dust correction used may subtly change the shape of the luminosity
function. Consequently, if a mass dependent dust correction \citep[as
  suggested in][]{Shimakawa15} is applied to the data instead of a
uniform dust correction for all objects the luminosity function will
flatten, increasing the number of HAEs with high H$\alpha$
luminosities. However, this will not affect the {\it excess} of bright
objects seen around radio galaxies compared to the field, unless there
is a different dust-stellar mass relation in these dense regions
compared to the field.

\begin{figure}
\centering
\includegraphics[width=0.95\columnwidth]{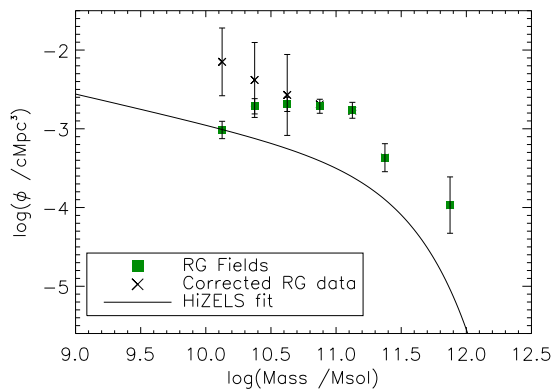}
\caption{Mass function of the H$\alpha$ emitters (HAEs) in the radio
  galaxy fields with the field line from the HiZELS survey
  \citep{Sobral13} at $z=2.23$. The raw values are shown as green
  boxes and the completeness-corrected values are shown as
  crosses. There is a clear excess of galaxies in the radio galaxy
  fields, which once completeness is corrected for, is consistent in
  shape with that of the HiZELS mass function, albeit offset by a
  factor of $\sim$ 3-5 (see Table \ref{passive}).}
\label{Massfn_com}
\end{figure}

The HAE mass function is shown in Fig.~\ref{Massfn_com}. This is again
compared to the field as determined from HiZELS \citep[smooth black
  line;][]{Sobral13}, which uses SED fitting to determine the mass of
the HAEs. We see an excess of galaxies compared to HiZELS, as
expected, that follows a similar shape to the HiZELS mass function at
high mass but again, the lowest mass bins are affected by
incompleteness. In order to correct the mass function without assuming
the distribution of HAEs in mass or luminosity-EW space from HiZELS,
we use Eq.~5 to estimate the mass of the HAEs as a function of
$Ks$-band magnitude assuming a constant correction factor (Eq.~6) of
-0.123 (i.e.~assuming $J-K=1$ - approximately the average colour of
the HAEs). The fraction of sources recovered from the reduced $Ks$
images at each magnitude was estimated by injecting circular,
Gaussian-profiled sources into the images, running SExtractor and
measuring the number recovered. From this the fraction of HAEs likely
to have been missed per mass-bin was estimated and the measured number
density increased by the inverse of this fraction. These corrections
were significant for the three lowest mass bins, ranging from 0.15 dex
for the third lowest to 0.9 dex for the lowest.  Having applied this
correction, the {\it shape} of the mass function in the radio galaxy
is consistent within the uncertainties with that of HAEs in the
general field. We note that previously \citet{Koyama13b} and
\citet{Cooke14} found excesses of line emitters in two radio galaxy
fields appeared to be confined to the most massive galaxies.

Thus, this work along with other studies with similar findings
\citep{Steidel05,Hatch11,Koyama13a,Cooke14} demonstrate that
protocluster fields, such as those around radio galaxies, contain an
excess of massive star-forming galaxies with comparatively high star
formation rates over those selected in the same manner from the same
volume (at the same redshift) in the field.

\subsection{HAE Properties}
The previous two sections have shown that the volume density of
star-forming galaxies is higher in the immediate environment of radio
galaxies than in the field. Once completeness corrections have been
applied, the shape of both the H$\alpha$ luminosity and stellar mass
functions for the line emitters derived in this work are consistent
with those of the field.
  
\begin{figure}
\centering
\includegraphics[width=0.95\columnwidth]{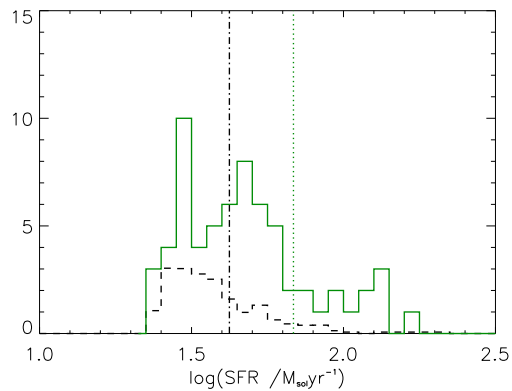}
\caption{Histogram of the H$\alpha$ star formation rates of HAEs in
  the radio galaxy fields and the HiZELS field (scaled to the same
  volume as the sum of the radio galaxy fields) with the mean of each
  distribution overplotted. Both samples have had the same HAE
  selection (from the NVSS J094748 field) applied. The mean star
  formation rate is higher in the radio galaxy fields, but a KS test
  cannot distinguish between the distributions at any significant
  level. }
\label{SFRdist}
\end{figure}

Fig.~\ref{SFRdist} shows the distribution of observed star formation
rates derived from H$\alpha$ for the radio galaxy fields and for the
HiZELS survey field when the same NVSS J094748 selection is applied to
both fields and the HiZELS values degraded to the same uncertainties
for a given flux/SFR as that of the radio galaxy field data. The mean
SFR is higher in the radio galaxy fields ($68\pm15$ versus $42\pm3$
M$_{\odot}$yr$^{-1}$ for HiZELS). A KS test on the two star formation
rate distributions cannot reject at any level of significance that
they are drawn from the same population, unsurprising given the
similarity in shape of the two H$\alpha$ luminosity functions once
incompleteness has been corrected for.

Despite the similarities in the shapes of the HAE line luminosity and
stellar mass functions for the radio galaxy fields and HiZELS once
corrected for incompleteness, we find more low equivalent width HAEs
in the radio galaxy fields than in HiZELS (see Fig.~\ref{EWdist}) (the
median EW for the radio galaxy and survey fields when an identical
selection is applied is 163$\pm$13 \AA\ and 120$\pm$40
respectively). The differences can in part be explained by the effect
of incompleteness on the lower mass HAEs - low mass, low equivalent
width HAEs are not detected and/or selected in both our observations,
and to a lesser extent in HiZELS, as they lie below the curved
selection line in colour-magnitude space (see
Fig.~\ref{MRC0200+015}). However, the lack of high mass {\it and} high
equivalent width HAEs in the radio galaxy fields and illustrated
  in the same figure is real - we see only one HAE with a NB
magnitude brighter than 20 and $K-NB > 1$ excluding the radio
galaxies. If the high mass objects had the same range of EW as for the
lower mass objects, they would have been selected.

The rest-frame EW measures the specific star formation rate (sSFR; the
SFR per unit stellar mass) of the galaxies.  For the objects meeting
our selection, the mean sSFR for the massive (M $>10^{10}$
M$_{\odot}$) HAEs is $\sim 1\times 10^{-9}$ yr$^{-1}$; around the
lower edge of the so-called main sequence of star formation at this
redshift \citep{Elbaz11,Karim11,Rodighiero14} and less at higher mass
($\sim 7\times10^{-10}$ yr$^{-1}$ at M $>10^{10.5}$ M$_{\odot}$ and
$\sim 4\times 10^{-10}$ yr$^{-1}$ at M $>10^{11}$
M$_{\odot}$). \citet{Elbaz11} derive a means sSFR of $\sim 2.5\times
10^{-9}$ for the main sequence of star forming galaxies at
$z=2.25$. In other words, the sSFR in these relatively strongly
star-forming galaxies, appears somewhat suppressed (by around 0.3-0.8
dex) relative to the main sequence of star formation at this
redshift. A similar result was found in \citet{Hatch11} in two
$z\sim2$ radio galaxy fields and in \citet{Cooke14} in a $z=2.5$ radio
galaxy field, although \citet{Cooke14} note that this difference goes
away when both samples are cut to stellar masses greater than
10$^{10}$ M$_{\odot}$. If so, the relatively low sSFR seen in these
massive galaxies could simply be related to mass (through downsizing
where more massive galaxies tend to form their stars earlier and
quicker than less massive galaxies) and not dependent on environment.

Given the evidence for completed red sequences in clusters at z
  $\sim1.5$ \citep[see e.g.][]{depropris15} and for very early
  completion of star formation in the most massive cluster galaxies
  \citep[e.g.][]{Mei06, Blakeslee03} present-day galaxies with
stellar masses comparable to those found in these overdensities appear
to form their stellar populations early ($z \gtrsim 2.5$; consistent
with stars in the most massive galaxies forming earlier than those in
the bulk of other galaxies) and over a short period of time, typically
less than 10$^9$ years \citep{Thomas10}, or a sSFR of
$>1$~Gyr$^{-1}$. Assuming the same timescale for the most massive HAEs
studied here, a minimum {\it average} SFR to build $10^{11}
  M_{\odot}$ over this time would be at least 100 M$_\odot$ yr$^{-1}$,
  similar to or larger than the measured sSFR of the HAEs in the radio
  galaxy fields. If, as is likely, star formation varied
stochastically  during formation, the bulk of the stellar
population will have formed during periods with significantly higher
sSFRs than the values observed here. Consequently, even though these
overdensities of HAEs are identified through ongoing significant star
formation, many of the galaxies with masses $>10^{10}$ M$_{\odot}$ are
likely to be past their peak in star formation {(the HAEs are
  observed to lie below the main sequence at this redshift), and
  therefore likely to be  in the process of quenching on their way to
becoming the passively evolving systems observed in the cores of
groups and clusters at lower redshifts.

\begin{figure}
\centering
\includegraphics[width=0.95\columnwidth]{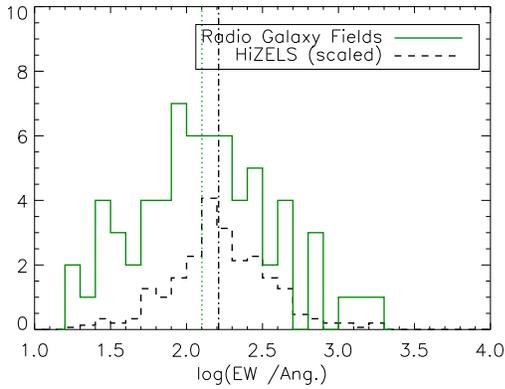}
\caption{Histogram of the H$\alpha$ equivalent width (EW) of HAEs in
  the radio galaxy fields and the HiZELS field (scaled to the same
  volume as the sum of the radio galaxy fields) with the median of
  each distribution overplotted. Again, both samples have had the same
  HAE selection (from the NVSS J094748 field) applied. We find more
  low EW HAEs in the radio galaxy fields compared to the HiZELS survey
  fields, largely due to the excess of high mass (and luminosity) HAEs
  in these fields. }
\label{EWdist}
\end{figure}

While this reduced star formation at this epoch may be a feature of
the evolution of massive galaxies in general it is worth exploring
whether, in the case of the galaxies in these fields, the presence of
a powerful radio galaxy in their immediate environment may be
affecting their ongoing star formation. If the radio galaxy is
affecting its local environment through heating of surrounding gas or
through direct ionization from the AGN we may expect that the
properties of the surrounding HAEs to change with distance from the
central radio galaxy. However, we find no trend of $Ks$ magnitude,
$Ks-$NB colour, EW or SFR with projected distance from the central
radio galaxy. We also find no trend of $Ks$ magnitude, $Ks-$NB
magnitude, EW or SFR with environmental density (calculated as the
number of HAEs within a 30 arcsec radius). Thus, there is no evidence
in this data of the radio galaxy affecting star formation in
neighbouring galaxies through proximity to that galaxy. This is
unsurprising as the observed fields (and therefore the scale length of
the overdensities) are much larger than the extent of the radio
emission from the radio galaxies.

\section{Conclusions}
We have studied the environment of seven $z=2.2$ radio galaxies with
broad and narrow-band imaging from VLT/HAWK-I designed to select
H$\alpha$ emitting galaxies (HAEs) at the radio galaxy redshifts. We
find that:
\begin{itemize}
\item All seven fields show a clear excess of HAEs relative to the
  expected surface density derived from field surveys. In particular,
  four of the seven fields are denser than 98 per cent of similar
  sized regions in the HiZELS survey.  One field in particular is very
  tightly clustered, the 1 arcmin$^2$ centred on the radio galaxy NVSS
  J094748 contains a density of HAEs so high that it is found only
  once over the same scale in the entire HiZELS survey. The fields of
  the other radio galaxies are overdense in HAEs spread across the
  wider HAWK-I field. The environments of the radio galaxies have
  properties consistent with those expected of the progenitors of rich
  groups and moderate mass clusters in the current day universe.
  Nevertheless, more richly clustered systems can be found in the
  $z\sim 2$ field \citep[e.g.][]{Geach12}. The {\it shapes} of the
  H$\alpha$ luminosity and HAE mass functions are indistinguishable
  from those of the field, the difference appears to be in their
  normalisation.

\item The excess of HAEs in the radio galaxy fields is evident across
  the range of the HAE mass function probed here, including high mass
  galaxies - indicative of significant prior growth of these
  systems. The median specific star formation rate for these massive
  (M $>10^{10}$ M$_{\odot}$) HAEs is $\sim 10^{-9}$ yr$^{-1}$ (around
  the lower edge of the main sequence of star formation at this
  redshift) and decreases with increasing mass. Given the timescale
  over which these galaxies form their stellar populations is expected
  to be less than a Gyr, these sources or their progenitors are likely
  to have previously been forming stars at higher rates than those
  observed.  Hence, these are massive galaxies undergoing (for them)
  moderate star formation at the observed epoch.

\item  There is no evidence of the star formation in individual galaxies
being affected by proximity to a radio galaxy based on a study of the
star forming parameters as a function of projected distance from the
radio galaxy.

\end{itemize}

\section{Acknowledgements}
We would like to thank the anonymous referee for their extremely
helpful comments that significantly improved this paper. We would also like to
thank Nina Hatch for helpful discussions about this work. KH
acknowledges funding from STFC. JPS acknowledges support from STFC
(ST/I001573/1) and a Hintze Fellowship. DNAM acknowledges support
through FONDECYT grant 3120214. This work is based on observations
made with ESO Telescopes at the La Silla and Paranal Observatory under
programme IDs 090.A-0387(ABC), 081.A-0932(A) and 083.A-0826(A). The
astronomical table manipulation and plotting software TOPCAT
\citep{Taylor05} was used in the analysis of these data. We use the publicly available HiZELS catalogues from \cite{Sobral13}.

\bibliographystyle{mn2e}
\bibliography{bibliography}

\end{document}